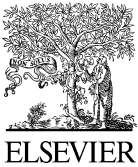

Available online at www.sciencedirect.com

ScienceDirect

Procedia Computer Science 00 (2019) 000–000

www.elsevier.com/locate/procedia

# 25th International Conference on Knowledge-Based and Intelligent Information & Engineering Systems

# Cybersecurity and Sustainable Development

Adam Sulich[a,♠], Małgorzata Rutkowska[b], Agnieszka Krawczyk-Jezierska[a,c,d], Jarosław Jezierski[b], Tomasz Zema[a]

[a] *Wroclaw University of Economics and Business, Faculty of Management, Komandorska 118-120, Wroclaw 53-345, Poland*
[b] *Wroclaw University of Science and Technology, Faculty of Computer Science and Management, Wyspianskiego 27, Wrocław 50-370, Poland*
[c] *The Center of Banking Law and Information, ul. Kruczkowskiego 8, 00-380 Warsaw, Poland*
[d] *The Warsaw Institute of Banking, ul. Solec 38/104; 00-394 Warsaw, Poland*

**Abstract**

Growing interdependencies between organizations lead them towards the creation of inter-organizational networks where cybersecurity and sustainable development have become one of the most important issues. The Environmental Goods and Services Sector (EGSS) is one of the fastest developing sectors of economy fueled by the growing relationships between network entities based on the ICT usage. In this sector the Green Cybersecurity is an emerging issue because it secures processes related directly and indirectly to the environmental management and protection. In the future the multidimensional development of the EGSS can help European Union to overcome the upcoming crises. At the same time computer technologies and cybersecurity can contribute to the implementation of the concept of sustainable development. The development of environmental technologies along with their cybersecurity is one of the aims of the realization of sustainable production and domestic security concepts among the EU countries. Hence, the aim of this article is a theoretical discussion and research on the relationships between cybersecurity and sustainable development in the inter-organizational networks. Therefore, the article is an attempt to give an answer to the question about the current state of the implementation of cybersecurity in relation to the EGSS part of the economy in different EU countries.

♠ corresponding author:
E-mail address: adam.sulich@ue.wroc.pl

## 1. Introduction

The Environmental Goods and Services Sector (EGSS) is one of the fastest developing areas of economy in the European Union [1,2]. This development is caused by the implementation of Sustainable Development Goals (SDGs) in the organization strategies [3,4] and by the growing importance of the digital technologies [1,5]. The EGSS is related to the Information and Communications Technology (ICT) sector, which is an extensional term for Information Technology (IT). In both sectors there are multiple strategies to achieve more sustainable development goals in organizations and on the level of the EU countries [6,7]. Surprisingly, the innovations introduced by the ICT sector gain more and more attention in the EGSS [2,8] and the green sectors of economy. Researchers [2,9] have discussed several approaches to using ICT in the service of sustainability, which have emerged during the last two decades in the academic and industrial spheres [10]. The ongoing Fourth Industrial Revolution (Industry 4.0) helps to increase automation and to improve communication, self-monitoring and the development of smart machines that can analyze and diagnose potential issues [11]. Industry 4.0 when combined with the circular economy, represents a new industrial paradigm enabling new natural resource strategies [12]. These approaches were then most often discussed in the context of the Jevons paradox [13,14], an economic argument implying that technological efficiency alone will not produce sustainability [2,15]. Researchers' consideration leads to the conclusion that a "combination of efficiency and sufficiency strategies is the most effective way to stimulate innovations which will unleash ICT's potential to support sustainability" [2]. The growing interest in digitization in almost all areas of modern life, and its dynamic development, especially now in the era of the pandemic, brings along both benefits and threats. Therefore, the current challenge for every user of ICT system is ensuring security in cyberspace in the area of bio-security which is also related to the achieving the SDGs. In the European Union (EU) it is important to combine internal and external security and the need to develop policies, actors and instruments that are consistent in this security context [16]. Both cybersecurity and Sustainable Development (SD) are the worldwide issues [17]. The European Union countries established goals dedicated both to the SD and cybersecurity, and they cooperate in various inter-organizational networks where cybersecurity and sustainable development have become an important issue. Increasing investments in Renewable Energy Sources (RES) can contribute to the SDGs, but also more often are dependent on the ICT systems and cybersecurity. On the other hand, these investments rely also on the banking industry. These complex relations have not been explored so far in the EGSS related literature. Therefore, this paper is an attempt to give an answer to the question about the current condition of the implementation of cybersecurity in EGSS and the related areas of economy in different EU countries.

The aim of this article is both a theoretical discussion and research on the relationship between cybersecurity and sustainable development in the inter-organizational networks. We understand the whole sector as an ecosystem whose relations can be measured by selected variables. The secondary data research presented the comparison between the EU countries on the level of the conditions created for such a unique combination visible in the EGSS. The purpose of this study is to analyze cyber threats in the aspect of environmental decision-making. Therefore, the decision-making problem discussed in this paper is also related to cyber threats.

The paper is organized as follows. After this brief introduction the literature review is provided and it covers the descriptive method followed by the statistical-analytical method. The descriptive method distinguishes and describes such specific phenomena as cybersecurity and sustainable development. The third part of this study is dedicated to the reference method description which backs the theoretical discussion. Then, a comparative analysis is used, which indicates the importance of the given issues when making decisions among the EU countries in the environmental aspect, including cybersecurity.

## 2. Literature Review

The Fourth Industrial Revolution [5], and thus the new emerging threats, have established new requirements for state security, such as control and protection of information in cyberspace [18]. These activities allow to counteract attacks by criminal groups and to prevent penetration by hostile entities [19]. Therefore, cybersecurity covers a set of issues related to providing protection in the area of cyberspace. The concept of cybersecurity is related, inter alia, to the protection of the information processing space and the interactions taking place in ICT networks [20,21]. The level of cybersecurity depends both on the degree to which the government mobilizes private entities to participate in security initiatives against common threats, as well as on the flexibility in relation to the expectations of public and private entities [19]. The first factor is commonly known as "compliance requirements" which are the base for cybersecurity provision, eg, in the financial sector. However, in many cases, they are not sufficient to answer all the challenges of technological development and the new threats induced by it. The ICT can have both negative and positive impacts on the environment [10]. On the one hand, the use of ICT has direct impact on electricity consumption, but on the other hand ICT can reduce emissions by building smarter cities, transportation systems, electrical grids and industrial processes [10]. The green approach in ICT recognizes this sector activities as a possible solution to many environmental and social problems [5].

Consequently, the EU considers achieving SD through the use of renewable energy as a way to mitigate the negative effects of climate change and to generate indirect economic benefits [9]. The Industry 4.0 relies on digital technologies such as wireless connectivity and sensors connecting everything to everyone to gather data and to analyze, visualize the entire production system [11]. Therefore cybersecurity gives inevitable support to innovative life cycle management, communication and development of smarter systems reconceptualizing waste as intrinsically valuable [1,11]. There is much hope that ICT can be a major part of the solution in dealing with climate change and related environmental challenges [10]. There is also a never-ending battle between security and profitability [22]. In turn, cyberspace is "the space for processing and exchanging information created by ICT systems, specified in Art.3 point 3 of the Act of February 17, 2005 on the computerization of the activities of entities performing public tasks along with the connections between them and relations with users"[23].

Technology helps us to save energy and resources, simultaneously, placing us at the front line in the war against growing in power cybercriminals. One of the important vectors of attack is cloud computing and widely used Data Sharing Networks (DNS). It is a convenient tool for low-cost data storage without heavy investment in infrastructure [24]. The major goal of cloud computing is to reduce the operating costs, increase both throughput and the reliability and availability [25]. However, cloud computing was ranked first among cybersecurity threats in 2020. To be effective in achieving goals business has to implement adequate security measures. This type of activity we call Green Cybersecurity, which also protects the processes related to the energy production, production of goods and services in EGSS. According to the Deloitte's report [26] "the interconnected nature of Industry 4.0-driven operations and the pace of digital transformation mean that cyberattacks can have far more extensive effects than ever before". Furthermore, the number of interconnected end-devices grows year by year and is forecast to reach 25.44 bln by 2030, as presented in Figure 1.

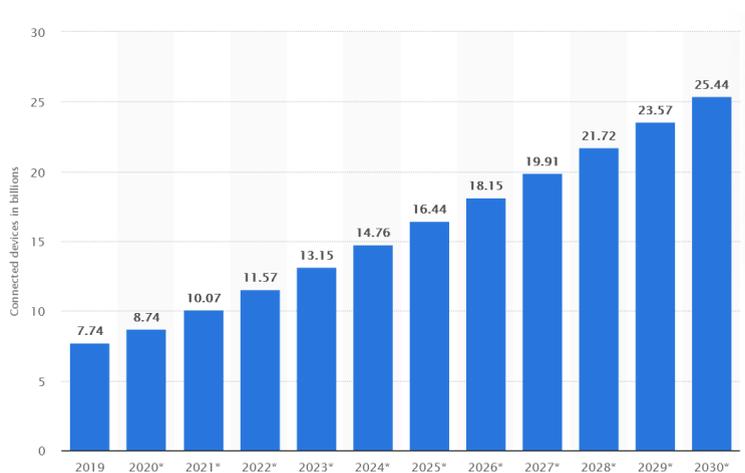

Figure 1. Number of Internet of Things (IoT) connected devices worldwide from 2019 to 2030 (in billions). Source: [27]

The Report predicts that the year 2021 will present the growth of the IoT devices up till 10.07 bln, whereas the Global Digital Report 2020 estimated that there were already 10.94 bln IoT devices in 2020 [28], see the Table 1. The biggest opportunities for ICT are in smart motor systems, logistics, buildings, grids and dematerialization [10]. It is important to understand the net impact of ICT on environment. "However, the empirical evidence on how ICT contributes to environmental sustainability particularly at macroeconomic and global level is scarce" [10].

Table 1. Connected Devices and the Internet of Things in bln

| Fixed phones | Mobile phones | PCS, laptops and tablets | Short range IoT devices | Wide area IoT devices |
|---|---|---|---|---|
| 1.36 | 7.94 | 1.57 | 10.94 | 1.88 |

Source: Own elaboration based on [28]

However, the back bone of the visible development is undoubtedly the financial industry. This sector is one of the most experienced by cyberattacks and the annual average cost of cybercrime was in 2017 the highest in this sector amounting to 18.28 mln USD annually per institution [29] and it was confirmed by Chris Thompson, Global Security and Resilience Lead – Financial Services in Accenture Security: "...the financial services industry continually has the highest cost of cybercrime" [30].

According to Kasperky Lab the largest number of devices was also in the banking industry, which is in the vanguard of technological changes in the world [31]. "The opportunities provided by technological advancements in banking are changing the industry as we know it, from an in-person experience to a personalized, AI-powered digital experience" [32]. Chat bots or intelligent analytical tools became common marks on the landscape of financial institutions. Undoubtedly, this is the factor which makes financial services the sector most vulnerable to cyber threats. The Sixth Annual Bank Survey by the US Conference of State Bank Supervisors presented the results of a survey according to which 70 per cent of the surveyed community banks ranked cybersecurity as the top concern for their institution [33]. Also 70 per cent of financial institutions experienced a security incident in the past year, according to 2019 "The Unknown Threat Report" by Clearswift [34]. Moreover, the UK Cybersecurity Company Carbon Black revealed the results of the survey among 25 CISOs (Chief Information Security Officers) in a report Modern Bank Heists 3.0 [35]. The results are presented in Table 2. It is worth noting that the COVID-19 pandemic has had an enormous impact on the cybersecurity of institutions and businesses bringing about the proliferation of cyberattacks on unprecedented scale.

Table 2. The results of the survey performed by Black Carbon in May 2020

| The area surveyed | Results in per cent |
|---|---|
| The per cent of financial institutions that reported an increase in cyberattacks over the past 12 months | 80 |
| The per cent of attacks that targeted either the healthcare sector or financial sector | 27 |
| The per cent of the increase of cyberattacks against the financial sector from February to April 2020 during the COVID-19 surge | 238 |
| The per cent of the surveyed financial institutions that said that cybercriminals had become more sophisticated | 82 |
| The per cent of surveyed financial institutions that reported increased attempts of wire fraud transfer | 64 |

Source: Own elaboration based on [35].

Furthermore, according to the same survey, the ransomware attacks against the financial sector increased by 9 times from the beginning of February to the end of April 2020, indicating again the correlation of the pandemic's surge and the growth of cyberattacks on financial institutions [35]. Financial industry is also ranked first when sensitive exposed files are concerned. According to the 2019 Data Risk Report by Varonis it had the highest number of exposed sensitive files from among all the examined industries, with 352,771 exposed sensitive files on average [36]. "In the face of technological advance and, as a result, the increasing threat of the loss of growing amount of data collected by financial institutions, it seems necessary to employ effective security measures in the process of information management" [37]. There is a growing number of financial institutions that implement ISMS (Information Security Management Systems) based on international standards such as i.a. ISO/EIC27001, not only to comply with regulatory requirements but also to make the security management more effective and, thus, saving costs of security breaches [38].

Financial institutions are becoming slowly part of the bigger business ecosystem and create inter-organizational networks due to digitalization process. This process is a derivative of the economic growth which indicates that a nation can produce more goods and services, and ICT enables goods and services to be produced more efficiently [10]. The extant research highlights the need for broadening the scope of the ICT value research beyond the established financial and economic frameworks. While there are regional studies on the potential of ICT for reducing emissions, this literature does not provide comparable empirical evidence on more international scale. There are reports which investigate the relationship between ICT (cybersecurity) and emissions for a sample of EU countries and "finds that ICT significantly decreases emissions per output in the energy supply sector" [10].

## 3. Method and Data Collection

For the evaluation of the Green Cybersecurity (GC), from among the methods of measuring the effects of green management or pro-ecological strategies of countries and local government units or companies, the reference method [39] or its modification is often used in the literature [40]. The reference method, called also the Hellwig's method [41], comes down to the determination of a synthetic variable being a function of the normalized features of the data input set. The method is also called the information capacity indicator method. The essence of this method lies in the procedure according to which, from the explanatory variables in the matrix, a combination of variables is selected. The so-called integral information capacity is the greatest. Moreover, this method allows to measure and compare variables of different sizes and dimensions with each other, because data standardization procedure is used [7,40].

The purpose of the method is to compare the level of the ICT usage and cybersecurity development described as the Green Cybersecurity (GC) at the international level and those creating a more sustainable economy in the EU [42]. The indicators were defined by Eurostat and their units are presented below (Table 1). The variables used in calculations were assigned by symbols x with the number lower index ($x_i$). As result, the total number of 7 variables was determined in this way.

**Table 3.** ICT usage in enterprises based on the Eurostat data

| Industry 4.0 area | Measured characteristic | Symbol |
|---|---|---|
| E-commerce | Enterprises having received orders online (at least 1%) - % of enterprises (tin00111) | $x_1$ |
| | Share of enterprises' turnover in e-commerce - % (tin00110) | $x_2$ |
| Connection to the internet | Enterprises with broadband access (tin00090) | $x_3$ |
| | Enterprises giving portable devices for a mobile connection to the internet to their employees (tin00125) | $x_4$ |
| E-business | Enterprises using radio frequency identification (RFID) instrument (tin00126) | $x_5$ |
| | Enterprises whose business processes are automatically linked to those of their suppliers and/or customers (tin00115) | $x_6$ |
| Competitiveness and innovation | Enterprises using software solutions, like CRM to analyse information about clients for marketing purposes (tin00116) | $x_7$ |

Source: Own study based on [18].

Secondary data from the year 2019 collected by Eurostat [18] were used for the calculations, which ensures the comparability and reliability of the data. The raw source data for each EU-28 country are enclosed in the appendices section (table A). The rational reason for the choice of the taxonomic method, especially the zero unitarization method [39] is the development of an econometric model of the Green Cybersecurity (GC) and the indication of its determinants in the further part of this paper. Moreover, the application of the standard method allows for the verification of the obtained results in comparison with the countries with similar development described in the literature. Since the set of independent features contains variables that cannot be aggregated directly using appropriate standardization, normalization formulas were applied. Among the formulas, the method of zero unitarization was selected to standardize the process based on the interval of a normalized variable. Variables that positively influence the described phenomenon are called stimulants. Their opposites are de-stimulants. Indicators are selected to a standardization process based on the following formulas [7,41]:

$$\text{for stimulants: } z_{ij} = \frac{x_{ij} - \min(x_{ij})_i}{\max(x_{ij})_i - \min(x_{ij})_i} \quad (1)$$

$$\text{for de-stimulants: } z_{ij} = \frac{\max(x_{ij})_i - x_{ij}}{\max(x_{ij})_i - \min(x_{ij})_i} \quad (2)$$

where:
$z_{ij}$ – is the normalized value of the *j*-th variable in the *i*-th country;
$x_{ij}$ – is the initial value of the *j*-th variable in the *i*-th country.

Diagnostic features normalized in an above-mentioned way take the value from the interval [0;1]. The closer the value to unity, the better the situation in terms of the investigated feature, and the closer the value to zero, the worse the situation.

In the next step, the normalized values of variables formed the basis for calculating the median and standard deviation for each of the countries studied. Median values were determined using the formula:

$$\text{for even number of observations: } Me_i = \frac{Z(\frac{m}{2})_i + Z(\frac{m}{2}+1)_i}{2} \quad (3)$$

$$\text{for odd number of observations: } Me_i = Z(\frac{m}{2}+1)_i \quad (4)$$

where:
$z_{i(j)}$ – is the *j*-th statistical ordinal for the vector $(Z_{i1}, Z_{i2}, \dots, Z_{im})$, i = 1, 2, …, n; j = 1,2, …, m.
In turn, the standard deviation was calculated according to the following formula:

$$S_{di} = \sqrt{\frac{1}{m}\sum_{j=1}^{m}(z_{ij} - \bar{z})} \quad (5)$$

Based on the median and standard deviation, an aggregate measure $w_i$ of the green management towards the GIR was calculated for each country:

$$w_i = M_{ei}(1 - S_{di}); w_i < 1 \quad (6)$$

Values of the measure close to one indicate a higher level of the greening of industrial parts of the economy in the specific member state, resulting in a higher rank. The aggregate measure prefers countries with a higher median of features describing the specific country and those with smaller differentiation between the values of features in the specific country expressed as standard deviation. The procedure chosen for evaluating the GIR provided a multidimensional comparative analysis. Such analysis allowed for a comparison between member states of the EU providing grounds for classifying them into uniform groups (Table 2).

Table 4. Green Growth Indicators aggregate measure comparative analysis

| Group | Mathematical characteristic | Meaning |
|---|---|---|
| I | $w_i \geq \bar{w} + S$ | high level |
| II | $\bar{w} + S > w_i \geq \bar{w}$ | medium-high level |
| III | $\bar{w} \geq w_i \geq \bar{w} - S$ | medium-low level |
| IV | $w_i < \bar{w} - S$ | low level |

Where $\bar{w}$ is the mean value of the synthetic measure; S is the standard deviation of the synthetic measure.

## 4. Results and Discussion

According to the $w_i$ values the EU countries were assigned to one of the groups concerning their level of evaluating the GC expressed in the greening economy.

**Table 5.** Groups of the EU countries with a similar level of ICT development related to the EGSS

| Group | Countries |
|---|---|
| I | Finland, Denmark, Sweden, Germany, Austria, Estonia, Latvia, United Kingdom, |
| II | Luxembourg, The Netherlands, Lithuania, Belgium, France, Czechia, Slovakia, Slovenia, Ireland |
| III | Poland, Hungary, Malta, Cyprus, Italy, Spain, Portugal, Greece |
| IV | Romania, Bulgaria, Croatia |

Source: Authors' calculations

The level of the GC was evaluated in the 28 EU countries based on 12 variables (Table 1) and the results of the analysis were presented in Table 3. The analysis shows that there are countries that operate in high-level management towards the GC and it presents countries with the best conditions for sustainable (green) development described at the international level.

## 5. Conclusions

The introduction of the ICT in every field, and the phenomena related to it such as the need for speed of operations and responsiveness, the need for more flexibility of organizations, the move to networking in the virtual domains, and reduction of wastes are the most important new developments that organizations have become familiar with. The literature shows multiple aspects of two main trends: cybersecurity and achieving SDGs. A recent literature review on Industry 4.0 presented that there is a connection between the SD and cybersecurity.

Automation of processes based on the cooperation of interconnected devices using Internet results in the enhanced exposure to cyber threats that can become the greatest drag chain to both ICT's development and EGSS. Hence, cybersecurity should be perceived as the indispensable element and should lie at the roots of the ongoing Green Technological Revolution. In the era of the development of new techniques and technologies, it should be noted that for investors, first of all, security is essential. Moreover, the development of information and communication technologies has, in practice, transformed every aspect of our lives today.

It should also be emphasized that cyberattacks are evolving and that is why it is so important to have measures (legal, regulatory and organizational) to control cybersecurity. Today, cybersecurity, data regulation and sustainability will be key to the digital transformation processes in the coming years. The banking and financial systems should engage and create useful tools for Sustainable Development

We have discussed several approaches to using ICT in the service of sustainability which have emerged during last decades. Therefore, the implementation of ICT in pro-ecological activities succeeds only when the enormous efficiency potential of these technologies is based on intelligent, informative, intensive and immaterial solutions and when it is the most important source of growth among the EU countries.

The ex-post analysis used in the study allowed for enriching the theoretical knowledge presented in the discussion part. On the other hand, the use of ex-ante analysis made it possible to determine the expected economic processes. The decision-making processes concerned the Green Cybersecurity and the aspect of Sustainable Development. The research was conducted to assess cybersecurity in the EU countries. The research results are presented in tables, enriched with graphic forms.

**Acknowledgments**

The project is financed by the Ministry of Science and Higher Education in Poland under the program "Regional Initiative of Excellence" 2019–2022, project number 015/RID/2018/19, total funding amount 10,721,040.00 PLN.

The project is financed by the National Science Centre in Poland under the programme "Business Ecosystem of the Environmental Goods and Services Sector in Poland" implemented in 2020-2022 project number 2019/33/N/HS4/02957 total funding amount 120,900.00 PLN.